# Competition of Superconductivity and Magnetism in $MoSr_2R_{1.5}Ce_{0.5}Cu_2O_{10-\delta}$ (R=Rare-earth, Mo-1222)

I. Felner and E. Galstyan

Racah Institute of Physics, The Hebrew University, Jerusalem, 91904, Israel.

We have investigated the $MoSr_2R_{1.5}Ce_{0.5}Cu_2O_{10-\delta}$ (Mo-1222R, R=rare earth) system by several complementary experimental techniques. In contrast to the iso-structural $RuSr_2R_{1.5}Ce_{0.5}Cu_2O_{10-\delta}$ (Ru-1222) system, in which superconductivity (SC) in the $CuO_2$ planes and *weak*-ferromagnetism in the Ru sub-lattice coexists, in Mo-1222, displays a competition between the two states, namely, SC vanishes when the magnetic order sets in. The contraction in the R elements leads to a change of the physical states. The light R ions (Pr and Nd) are paramagnetic down to 5 K, whereas the middle R ions (Sm and Eu) are SC at $T_C$ 18-23 K respectively. The SC charge carriers originate from the *$CuO_2$ planes,* and annealing under oxygen pressures does not affect $T_C$. A simple model for the SC state is proposed. For the heavy R elements Ho-Lu and Y, the pentavalent *Mo layers* are antiferromagnetically (AFM) ordered at $T_N$ ranging from 13-26 K. For R=Gd, the sample is not SC and exhibit two magnetic transitions at 11 and 184 K. Both the SC or AFM states depend strongly on the R/Ce ratio and for R/Ce=1, both states are suppressed.

Introduction

Superconductivity (SC) and ferromagnetism (FM) are generally believed to be mutually antagonistic states. Coexistence of weak-ferromagnetism (W-FM) and SC was discovered a few years ago in $RuSr_2R_{2-x}Ce_xCu_2O_{10}$ (R= Eu and Gd, Ru-1222) layered cuprate systems[1-2], and more recently[3] in $RuSr_2GdCu_2O_8$ (Ru-1212). The SC charge carriers originate from the $CuO_2$ planes and the W-FM state is confined to the Ru layers. In both systems, the states *coexist* intrinsically on the microscopic scale. The Ru-1222 materials display a magnetic transition at $T_M$= 125-180 K and bulk SC below $T_C$ = 32-50 K ($T_M > T_c$) depending on oxygen concentrations and sample preparation[1]. The hole doping of the Cu-O planes, can be optimized with appropriate variation of the R/Ce ratio[4]. X-ray-absorption spectroscopy reveals that the Ru ions are $Ru^{5+}$ irrespective of the Ce concentration [5]. The magnetic state of the Ru sublattice is not affected by the presence or absence the SC State, indicating that the two states are practically decoupled [6]. It is also apparent, that bulk SC *only*, appears in the M-1222 (M= Nb and Ta) with $T_C$~28-30 K, in which the M ions are pentavalent[7].

During the course of studying the Ru-1222 system, we noticed that Ru can be replaced completely by Mo ions and that the $MoSr_2R_{1.5}Ce_{0.5}Cu_2O_{10}$ (Mo-1222) system is iso-structural to the Ru-1222 one. The Mo-1222 system can be obtained with most of the R elements (Pr-Yb and Y). In contrast to the M-1222 (M=Nb, Ta) systems, which are SC only and to the Ru-1222 materials, described above, in Mo-1222, the SC and the magnetic states compete with each other. Materials, which are SC, are not AFM, and vice versa. The ionic radii of the R ions determine whether the material is paramagnetic (PM), SC or AFM ordered. We show here that in Mo-1222: (a) the light R element (Pr and Nd) are PM. (b) the middle R=Sm and Eu samples are SC at $T_C$= 18-23 K, and (c) that the heavy R ions (Ho-Yb and Y) are AFM ordered with $T_N$= 13-26 K. We also present the magnetic curves of $MoSr_2Gd_{1.5}Ce_{0.5}Cu_2O_{10}$ in which show two magnetic transitions at 11 and 184 K, and construct the SC-AFM phase diagram in Mo-1222.

**Experimental details**

Ceramic samples with nominal composition $MoSr_2R_{1.5}Ce_{0.5}Cu_2O_{10}$ (Mo-1222R) and $MoSr_2RCeCu_2O_{10}$ (R/Ce=1) were prepared by a solid-state reaction technique. Prescribed amounts of $R_2O_3$, $CeO_2$, $SrCO_3$, Mo, and CuO were mixed and pressed into pellets and preheated at 950° C for 1 day. The products were cooled, reground and sintered at 1050° C for 2 days under oxygen atmosphere and then furnace cooled. Part of the asp Mo1222R (R= Eu, Gd and Y) samples were re-heated for 12 h at 800° C under high oxygen pressure (70 atm.).

Powder X-ray diffraction (XRD) measurements indicate that all samples are nearly single phase (~96%) materials and confirmed the tetragonal structure. The XRD patterns contain a few minor additional peaks and all attempts to completely get rid of them were unsuccessful. For the R=La and Lu samples, the extra peak intensity exceed 25%, therefore their lattice parameters are not listed. Within the instrumental accuracy, the lattice parameters of R/Ce=1 samples are similar to the corresponding Mo-1222R compounds. The microstructure and the phase integrity of the materials were investigated by QUANTA (Fri Company) scanning electron microscopy (SEM) and by a Genesis energy dispersive x-ray analysis (EDAX) device attached to the SEM. Zero-field-cooled (ZFC) and field-cooled (FC) dc and ac magnetic measurements in the range of 2-400 K were performed as described in Ref. 4

**Experimental results and discussion**

Least squares fits of the XRD patterns of the $MoSr_2R_{1.5}Ce_{0.5}Cu_2O_{10}$ compounds on the basis of a tetragonal structure (SG I4/mmm) left a few minor reflections most of them belonging to $R_2O_3$ and to the $SrMoO_4$ phase and yield the lattice parameters given in Table 1. The variation of the *a* lattice constant shown in Fig. 1, is attributed to the lanthanide contraction of the $R^{3+}$ ions. The *c* constant can be considered as remaining constant. The similarity of *a* for R=Pr and Nd and the excess of *a* for R=Yb (Fig. 2), are most probably due to the mixed valence states of Pr and Yb. The morphology detected by the SEM, shows a smooth and uniform surface for the granular the Mo-1222Eu sample, with typical grain size of 2-3 μm. For Mo-1222Y the grain size is not well defined. EDAX analysis confirms the initial stoichiometric composition of R:Ce:Sr and Cu, whereas a deficiency in the Mo content is obtained due to its volatility. We also detected a few separate sphere grains of the $SrMoO_4$ phase.

**(a) Paramagnetism in $MoSr_2R_{1.5}Ce_{0.5}Cu_2O_{10}$ (R=Pr and Nd).** The dc magnetic susceptibility $\chi$(T) (=M(T)/H) curves of Mo-1222R (R=Pr Nd and La), exhibit normal PM behavior down to 5 K and the isothermal M(H) (up to 5 T) curves are linear. The $\chi$(T) curves can be fitted by the Curie-Weiss (CW) law: $\chi = \chi_0 + C/(T-\theta)$, where $\chi_0$ is the temperature independent part of $\chi$, C is the Curie constant, and $\theta$ is the CW temperature. $SrMoO_4$ detected by EDAX is Pauli-paramagnetic[8] and does not contribute to C. In order to get the net R contribution to C, we measured the $\chi$(T) of Mo-1222Y (see below) and $\chi$(T) of $YBa_2Cu_2O_7$ (at T>100 K) which is roughly temperature independent (1.8-2*$10^{-4}$ emu/mol Oe). After subtracting the two contributions (2/3$\chi$(T) of Y123) from the measured $\chi$(T), we obtained C and the $P_{eff}$ values listed in Table 1. [The same procedure was done for all C values given in Table 1] For the PM Mo-1222R (R=Pr and Nd) materials, the extracted $P_{eff}$ =2.85 and 2.31 $\mu_B$ respectively, are lower than the 3.56 and 3.62 $\mu_B$ expected for $Pr^{3+}$ and $Nd^{3+}$. This reduction is probably due to strong crystal field effects and/or to a mixed valence state of Pr as stated above. Note the difference between the two negative θ values of the materials.

**(b) Superconductivity in the $MoSr_2R_{1.5}Ce_{0.5}Cu_2O_{10}$ (R=Sm and Eu).** ZFC and FC magnetic curves for Mo-1222Eu are presented in Fig. 2. The onset of SC at $T_C$= 23 K was also obtained by ac measurements. Annealing under high oxygen pressure, did not change the $T_C$ value. On the other hand, increasing the Ce concentration suppresses SC and $MoSr_2EuCeCu_2O_{10}$ is PM down to 5 K (Fig. 2). For R=Sm, $T_C$= 18 is obtained. Above $T_C$, for R=Sm, the χ(T) plot, follows the CW law (Table 1), whereas, for R=Eu, the χ(T) curve is linear and does not obey the CW law.

With the purpose of acquiring information about the critical current density $J_C$, we have measured at 5 K the magnetic hysteresis (Fig. 2). Following the Bean's approach $J_C(H)$=30 ΔM/d, where ΔM is the difference in the M at the same H, and d=2.5 μm, we obtained: $J_C$= 4.4*10$^4$ A/cm$^2$ (at H=0) a value which compares well with $J_C$ obtained in Ru-1222 under the same conditions[9].

Our intuitive explanation as to why all the M-1222 materials ($M^{5+}$) are SC is discussed in Ref. 4. Assuming that Mo is pentavalent, we may apply the same model to the Mo-1222 materials. In the well-established phase diagram for $La_{2-x}Sr_xCuO_4$, the parent $La_2CuO_4$ is AFM and insulating and replacing $Sr^{2+}$ for $La^{3+}$ varies the hole density *p*. In Mo-1222 ($Sr^{2+}$, $Cu^{2+}$ and $O^{2-}$), Ce is tetravalent[5] and our Mossbauer spectroscopy (MS) study performed at RT on $^{151}$Eu, shows a singlet with an isomer shift of 0.04(2) mm/s (relative to $Eu_2O_3$), indicating that the Eu is trivalent. Given that Mo is pentavalent, a straightforward valence count yields a fixed oxygen concentration of 10. Thus, we argue that the $MoSr_2EuCeCu_2O_{10}$ sample serves as the parent compounds (similar to $La_2CuO_4$). Hole doping of the Cu-O planes, which results in SC, can be achieved with increasing of the $R^{3+}/Ce^{4+}$ ratio ($R^{3+}$ ions are replaced for $Ce^{4+}$) and indeed, $MoEu_{1.5}Ce_{0.5}Sr_2Cu_2O_{10-\delta}$ is SC. The optimal doping of Ce is now under investigation.

**(C) Antiferromagnetism in the $MoSr_2R_{1.5}Ce_{0.5}Cu_2O_{10}$ (R=Ho-Lu and Y).** The ZFC and FC curves of Mo-1222R (R=Y, Er and Tm) measured at 5-10 Oe, are presented in Fig. 3. One definitely sees the irreversibility and the peaks in the ZFC plots, typical of AFM ordering. We define $T_N$ (Table 1) as the merging point of these branches or, alternatively, as the inflection point in the dχ(T)/dT of the FC curve. Similar behavior was observed for R=Ho and Lu. For R=Yb, no irreversibility is obtained and $T_N$ =16 K was defined as the kink observed in both ZFC and FC curves. All M(H) curves measured below and above $T_N$ are linear up to 5 T and no hystersis is observed.

Above $T_N$, the χ(T) curve for Mo-1222Y follows the CW law, and provides useful information on the Mo valence. Taking into account the existence (of about 5-6%) of the $SrMoO_4$ phase and by subtracting the Cu contribution (see above) we obtained $P_{eff}$ =1.68 $\mu_B$ a value which is in good agreement with 1.73 $\mu_B$ expected for $Mo^{5+}$ ($4d^1$, S=0.5). Therefore, we argue with high confidence, that the prominent feature shown in Fig. 3, as well as the SC state described above, are related to the $Mo^{5+}$ sublattice. This preferable interpretation, which invokes analogy to the Ru-1222 system, means that the Mo layers are AFM ordered at relatively low temperatures (13-26 K). The $P_{eff}$ values for R=Ho-Yb (Table 1) are in good agreement with their calculated $R^{3+}$ free ion values. Note the negative θ obtained which is consistent with an AFM order.

For R=Tb and Dy, the χ(T) curves have a PM-like behavior down to 5 K. For R= Dy, the small contribution of the Mo AFM signal, (0.2 emu/mol Oe for R=Y see Fig. 5 inset), is probably masked by the high PM susceptibility of $Dy^{3+}$. The $P_{eff}$ value obtained fits well the expected value of $Dy^3$. For R=Tb, $P_{eff}$= 8.7 $\mu_B$ deduced, is smaller than the 9.72 $\mu_B$ expected for free ion $Tb^{3+}$, suggesting that Tb has an intermediate valence-state, which also affects its magnetic behavior. This is reminiscent of the exception of the non-SC Tb123, among all other R123 materials.

**(D) The $MoSr_2Gd_{1.5}Ce_{0.5}Cu_2O_{10}$ sample.** This sample shows two magnetic transitions at 11 K and at 184 K and positive θ (Fig. 4). The M(H) curves are linear and no hystersis is observed. At 5 K, for H> 3 T a tendency toward PM saturation (of Gd ions) is obtained. No magnetic transition is observed down to 1.8 K (Fig 4 inset). This is in contrast to Gd123, M-1212 (M=Ru and Mo) and Ru-1222 in which Gd is AFM ordered at 2.2-2.6 K. (I) We may suggest that the peak at 11 K is related to the Gd sublattice and the highest transition is due to the Mo layers. (II) A more preferable interpretation is that both transitions are related to the Mo sublattice, whereas the origin of the second one is not known yet. Our supporting evidence is the fact, that in $MoSr_2GdCeCu_2O_{10}$ (Gd/Ce=1) both anomalies are absent (Fig. 4), suggesting that the two magnetic transitions are connected to each other. The PM values extracted for $MoSr_2GdCeCu_2O_{10}$ are:θ= -4.4 K and $P_{eff}$ = 7.85 $\mu_B$/Gd which fits well the theoretical value of 7.94 $\mu_B$/Gd. At the present moment we cannot explain as to why the R=Gd sample behaves so differently from all the rest of the heavy Mo-1222R compounds.

**In conclusion**, we demonstrate that the behavior of Mo-1222 system is entirely different from $Ru^{5+}$-1222 one. In Mo-1222 ($Mo^{5+}$), the physical state depends strongly on the ionic radii of the R ionst. For large R ions, the samples are PM (Fig. 1). Once the *a* lattice parameter is contracted, SC in the $CuO_2$ planes is induced at $T_C$ ~ 20 K. Further contraction of *a* leads to an AFM order (at $T_N$ 13-26 K) in the Mo-O layers and to suppression of SC. The SC and AFM (which compete each other) states depend on the R/Ce ratio and both disappear for R/Ce=1. The magnetic behavior of the Gd sample is different. Neutron diffraction as well as MS on $^{155}$Gd studies, are now being carried out to determine the magnetic structure of this system.

**Acknowledgments** We are grateful to Prof. I. Nowik for the Mossbauer data. This research was supported by the Israel Academy of Science and Technology and by the Klachky Foundation for Superconductivity.

Table1. Lattice parameters and magnetic parameters of MoSr$_2$R$_{1.5}$Ce$_{0.5}$Cu$_2$O$_{10}$ (ND=not determined )

| R | a (Å) | c (Å) | T$_C$/T$_N$ (K) | C (emu*K /mol*Oe) | θ (K) | 10$^3$ χ$_0$ (emu/ mol*Oe | P$_{eff}$ (μ$_B$) |
|---|---|---|---|---|---|---|---|
| Pr | 3.861(1) | 28.38(1) | PM | 1.51(2) | -14.1(2) | 1.1(1) | 2.85 |
| Nd | 3.860 | 28.52 | PM | 1.00 | -2.1 | 4.5 | 2.31 |
| Sm | 3.848 | 28.43 | T$_C$=18(1) | 0.12 | -12.4 | 1.91 | 0.80 |
| Eu | 3.838 | 28.41 | T$_C$=23(1) | Not CW | | | |
| Gd | 3.833 | 28.43 | 15, 184 | 9.6 | 14.7 | 3.0 | 7.16 |
| Tb | 3.818 | 28.33 | ND | 14.2 | -8.5 | 3.0 | 8.70 |
| Dy | 3.815 | 28.37 | ND | 19.4 | -4.9 | 11.0 | 10.1 |
| Ho | 3.812 | 28.40 | 13(1) | 21.4 | -4.4 | 4.7 | 10.6 |
| Er | 3.810 | 28.44 | 26 | 19.6 | -11.3 | 0.1 | 10.1 |
| Tm | 3.797 | 28.30 | 13 | 10.7 | -22.0 | 1.84 | 7.56 |
| Yb | 3.807 | 28.41 | 16 | 2.38 | -7.5 | 2.2 | 3.58 |
| Y | 3.812 | 28.43 | 13 | 0.35 | -1.4 | 0.5 | 1.68 |

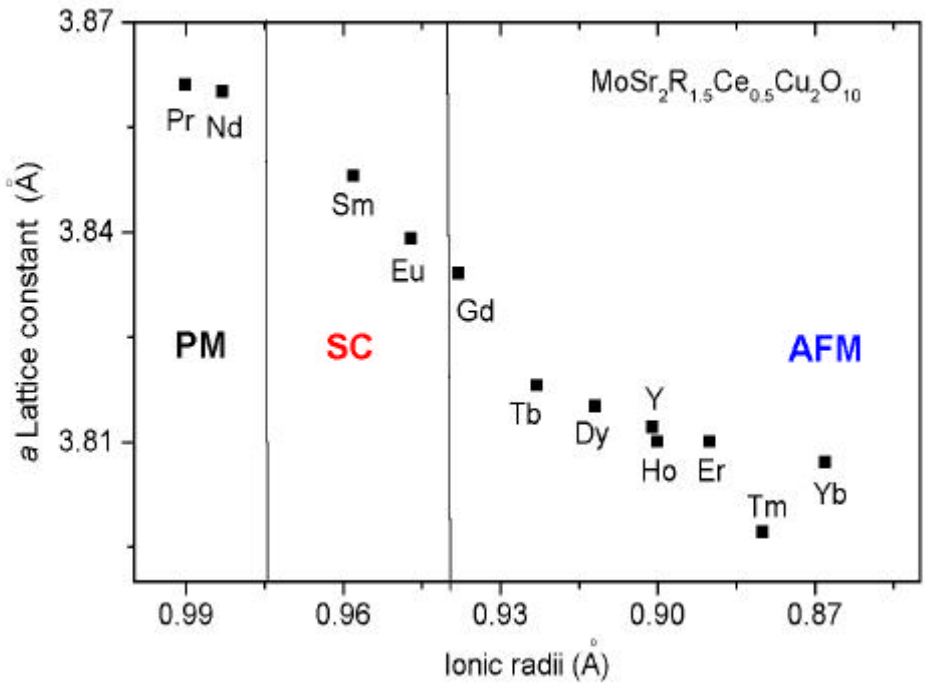

Fig. 1. The phase diagram and the *a* lattice parameter as a function of the ionic radii in Mo-1222.

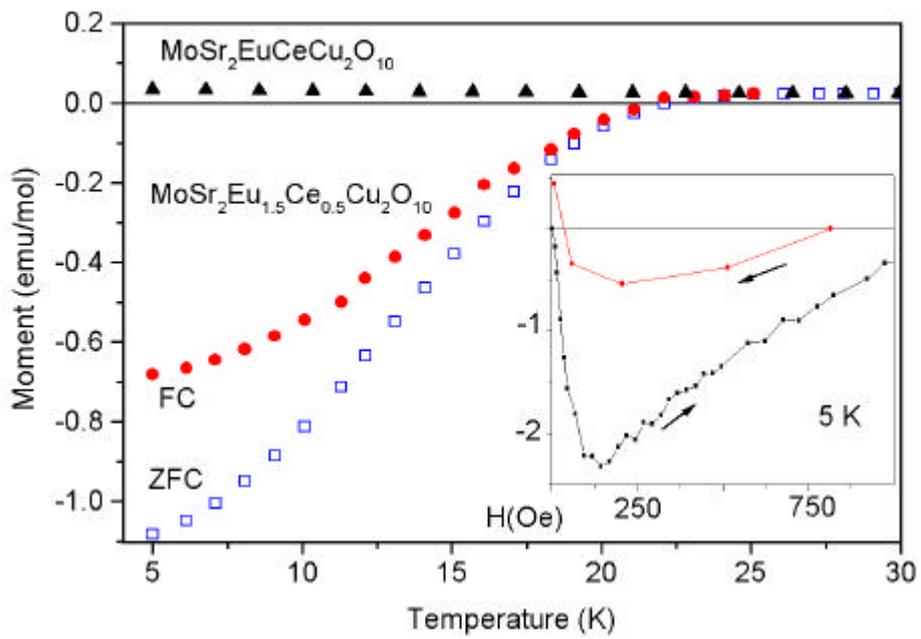

Fig. 2 Magnetic moment of MoSr$_2$Eu$_{1.5}$Ce$_{0.5}$Cu$_2$O$_{10}$ (SC) and MoSr$_2$EuCeCu$_2$O$_{10}$ measured at 3 Oe. The inset shows the hystersis loop at 5 for the SC material.

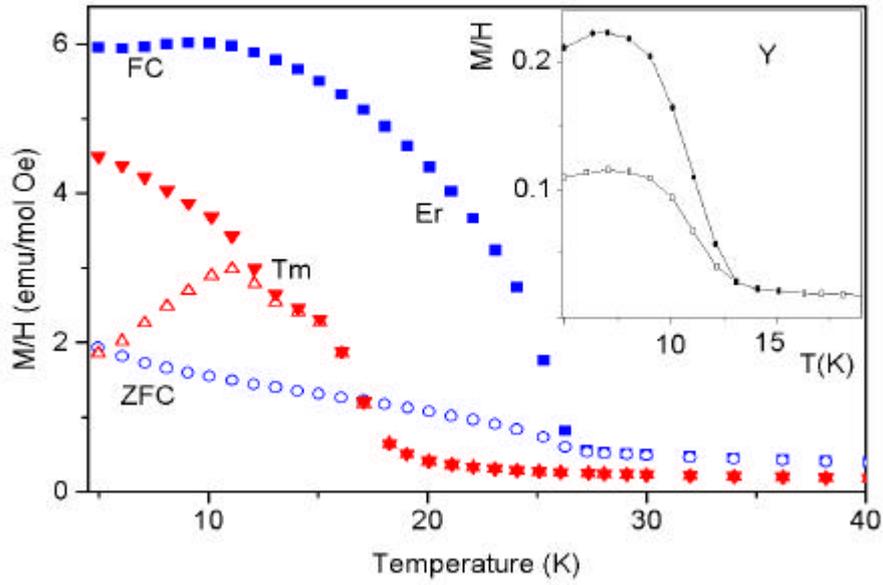

Fig. 3. ZFC and FC susceptibility curves measured at low applied fields of Mo-1222R R= Er, Tm and Y (inset).

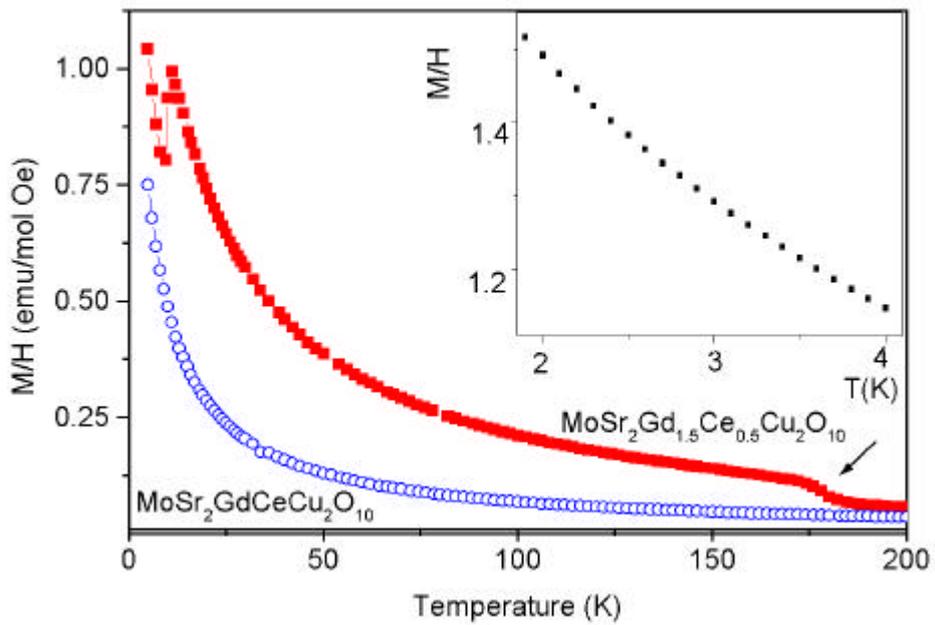

Fig. 4. M/H of $MoSr_2Gd_{1.5}Ce_{0.5}Cu_2O_{10}$ and $MoSr_2GdCeCu_2O_{10}$ measured at 20 Oe. The inset shows the M/H curve for $MoSr_2Gd_{1.5}Ce_{0.5}Cu_2O_{10}$ at T<4 K.